\documentclass[proceedings, preprint]{rmaa}

\usepackage{paralist}

\usepackage{psfrag,color}

\SetYear{2011}
\SetConfTitle{Fourth Science with the GTC Meeting}

\title{SHARDS: Survey for High-z Absorption Red \& Dead Sources} 

\author{
  Pablo G. P\'erez-Gonz\'alez,\altaffilmark{1,2} 
  Antonio Cava,\altaffilmark{1}
  and the SHARDS Team\altaffilmark{3}}

\altaffiltext{1}{Departamento de Astrof\'{\i}sica, Facultad de CC.
  F\'{\i}sicas, Universidad Complutense de Madrid, E-28040 Madrid,
  Spain}

\altaffiltext{2}{Associate Astronomer at Steward Observatory, UofA}

\altaffiltext{3}{http://guaix.fis.ucm.es/$\sim$pgperez/SHARDS/team.html}

\shortauthor{SHARDS Team}
\shorttitle{Update on SHARDS}

\listofauthors{P\'erez-Gonz\'alez et al.}
\indexauthor{P\'erez-Gonz\'alez, P. G.}
\indexauthor{Cava, A.}
\indexauthor{Team, SHARDS}

\abstract{

  SHARDS, an ESO/GTC Large Program, is an ultra-deep (26.5~mag)
  spectro-photometric survey with GTC/OSIRIS designed to select and
  study massive passively evolving galaxies at z$=$1.0--2.3 in the
  GOODS-N field using a set of 24 medium-band filters
  (FWHM$\sim$17~nm) covering the 500--950~nm spectral range.  Our
  observing strategy has been planned to detect, for z$>$1 sources,
  the prominent Mg absorption feature (at rest-frame $\sim$280~nm), a
  distinctive, necessary, and sufficient feature of evolved stellar
  populations (older than 0.5~Gyr). These observations are being used
  to: (1) derive for the first time an unbiased sample of high-z
  quiescent galaxies, which extends to fainter magnitudes the samples
  selected with color techniques and spectroscopic surveys; (2) derive
  accurate ages and stellar masses based on robust measurements of
  spectral features such as the Mg$_\mathrm{UV}$ or D(4000) indices;
  (3) measure their redshift with an accuracy $\Delta$z/(1+z)$<$0.02;
  and (4) study emission-line galaxies (starbursts and AGN) up to very
  high redshifts. The well-sampled optical SEDs provided by SHARDS for
  all sources in the GOODS-N field are a valuable complement for
  current and future surveys carried out with other telescopes (e.g.,
  {\it Spitzer}, HST, and {\it Herschel}).

}

\resumen{

  SHARDS (siglas en ingl\'es de la Exploraci\'on de galaxias con
  absorci\'on rojas y muertas a alto desplazamiento al rojo), un
  ESO/GTC Large Program, es una exploraci\'on espectro-fotom\'etrica
  super-profunda (26.5~mag) realizada con GTC/OSIRIS y dise\~nada para
  seleccionar y estudiar galaxias masivas evolucionando pasivamente a
  z$=$1.0--2.3 usando un juego de 24 filtros de anchura media
  (FWHM$\sim$17~nm) a 500--950~nm y apuntando a GOODS-N. Nuestra
  estrategia observacional ha sido ideada para detectar la marca
  espectral conocida como absorci\'on de Mg (a $\sim$280~nm), que
  constituye una caracter\'{\i}stica distintiva, necesaria y
  suficiente de las poblaciones estelares evoulcionadas (m\'as viejas
  que 0.5~Gyr). Nuestras observaciones est\'an siendo analizadas para:
  (1) construir por primera vez una muestra no sesgada de galaxias
  quiescentes a alto-z, llegando a magnitudes m\'as d\'ebiles de lo
  conseguido hasta ahora con t\'ecnicas espectrosc\'opicas y
  fotom\'etricas; (2) derivar edades y masas estelares precisas
  bas\'andonos en medidas robustas de \'{\i}ndices espectrales como el
  de Mg$_\mathrm{UV}$ o D(4000); (3) estimar desplazamientos al rojo
  con una precisi\'on de $\Delta$z/(1+z)$<$0.02; y (4) estudiar
  galaxias con l\'{\i}neas de emisi\'on (con formaci\'on estelar o
  AGN) hasta desplazamientos al rojo altos. Adem\'as, las SEDs
  \'opticas de resoluci\'on intermedia proporcionadas por SHARDS
  ser\'an de gran utilidad para otras exploraciones cosmol\'ogicas
  llevadas a cabo con otros telescopios (p.e., HST, {\it Spitzer} y
  {\it Herschel}).

}

\addkeyword{galaxies: photometry}
\addkeyword{galaxies: quiescent}
\addkeyword{galaxies: starburst}
\addkeyword{galaxies: high-redshift}

\begin{document}
\maketitle

\section{Introduction}
\label{sec:intro}

\hspace{0.5cm} One of the most interesting results in Extragalactic
Astronomy in the last decade is the discovery of a numerous population
of massive galaxies
($\mathcal{M}$$\gtrsim$10$^{11}$~$\mathcal{M}_\odot$) at high redshift
(e.g., Franx et al. 2003). Some of them are already evolving passively
(Daddi et al. 2004), being good candidates for the progenitors of
massive nearby ellipticals (Hopkins et al. 2009). Even more
puzzlingly, these galaxies present very small sizes, and thus large
mass densities (Daddi et al. 2005, Trujillo et al. 2007, Toft et al.
2007) comparable to the density of a globular cluster (Buitrago et al.
2008). The existence of very compact massive dead galaxies at
high-redshift is extremely challenging for models of galaxy formation,
based on the hierarchical $\Lambda$CDM paradigm (e.g, Baugh et al.
1996, Cole et al. 2000, de Lucia et al. 2006, Croton et al. 2006).

In addition, several works have found compelling evidence of a high
formation redshift (z$\gtrsim$2) jointly with a high star formation
efficiency for the most massive galaxies (Cimatti et al. 2006; Bundy
et al. 2007; P\'erez-Gonz\'alez et al. 2008a; Ilbert et al. 2009), a
scenario known as {\it downsizing} (Cowie et al. 1996).  Furthermore,
other studies report the detection of a red sequence up to z$\sim$2,
populated by post-starburst galaxies (Labb\'e et al. 2007; Arnouts et
al. 2007; Kriek et al. 2008; Fontana et al. 2009). These works have
shown that a key epoch for the study of the formation of the red
sequence is 1$<$z$<$2.

The SHARDS ESO/GTC Large Program aims at building an unbiased sample
of quiescent ETGs at z$>$1 and providing with reliable observational
estimations of relevant quantities such as their number density,
stellar masses, ages, and sizes. Reproducing these results at high-z
constitutes a major tests for the predictions of galaxy formation
models (de Lucia et al. 2006, Croton et al. 2006, Hopkins et al.
2008). Therefore, our Large Program will be a significant step forward
in our understanding of galaxy formation, shedding light into the
early massive star formation events and quenching mechanisms.

In this conference proceedings, we describe the observational strategy
of the survey and present the first results obtained by SHARDS
concerning the study of emission-line galaxies at various redshifts,
and the characterization of red and dead sources at z$>$1. More
information about the study of emission-line galaxies with SHARDS can
be found in other conference proceedings in this volume (Cava et al.
2012, Rodr\'{i}guez-Espinosa et al. 2012).

\vspace{-0.2cm}
\section{Survey description}
\vspace{-0.1cm}

SHARDS is using OSIRIS on the Spanish 10.4m telescope, GTC, to obtain
medium-band imaging through 24 filters (FWHM$\sim$17~nm), and covering
the GOODS-N field down to the 26.5~mag (AB) in each filter. Our goal
is to probe the rest-frame UV spectral range (250-350~nm) for galaxies
at z$>$1.0, counting with enough spectral resolution (R$\sim$50) to
detect and accurately measure the Mg$_\mathrm{UV}$ absorption. This is
a distinctive, necessary and sufficient feature of massive quiescent
ETGs at high redshift. A positive detection of the feature with our
spectro-photometric technique will confirm that the galaxies are
quiescent and provide us with an estimate of the age of the last star
formation burst (Daddi et al. 2005) and the galaxy redshift (with an
accuracy $\Delta$z/(1+z)$<$0.02). OSIRIS provides imaging capabilities
up to $\lambda$$\sim$950~nm, which means that we can probe the Mg
absorption at 1.0$<$z$<$2.3.

The resolution of this observing strategy is similar to that used by
the Grism ACS Program for Extragalactic Science (GRAPES; Daddi et al.
2005) and the Probing Evolution And Reionization Spectroscopically
project (PEARS; PI: Malhotra) to study quiescent ETGs through HST
grism slitless spectroscopy. Our project covers 10 times more area
than the one used in Daddi et al. (who used the UDF), and reach
1.5~mag fainter.

SHARDS has completed 75\% of the total 180 hours of observing time
awarded to the project, as of January 2012. The data quality is
excellent, reaching fainter fluxes than expected ($\sim$27~mag) and
virtually all data present sub-arcsecond seeing.

\section{Emission-line galaxies in SHARDS}
\label{sec:elgs}
\vspace{-0.1cm}

The comparison of fluxes measured in narrow-band images with those
from broad-band data in the same spectral region is a widely used
technique to select and study galaxies with emission-lines at low- and
intermediate- (e.g., Villar et al.  2008), and high-redshifts (e.g.,
Ouchi et al. 2008). The technique is based on the detection of a
prominent color when an emission-line of a galaxy (at an appropriate
redshift) lies within the narrow-band filter pass-band, in comparison
with a broad-band filter, whose measurement is dominated by the
continuum around that line.

\begin{figure}[!h]
  \includegraphics[width=7.1cm,angle=-90]{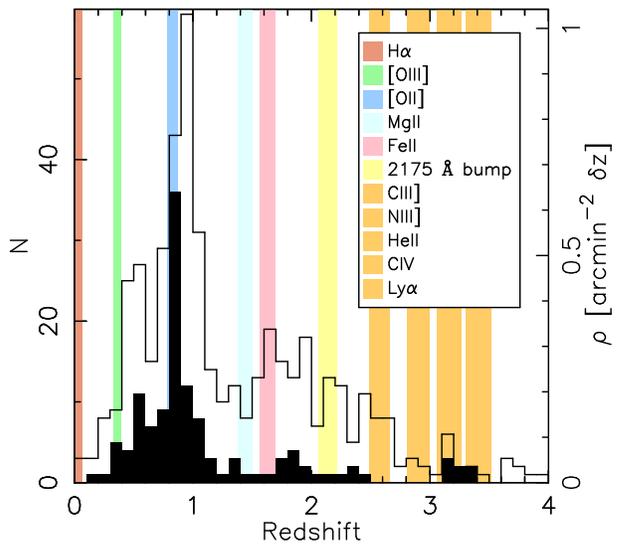}
  \caption{Redshift distribution of the galaxies selected as
    emission-line candidates using the SHARDS data at 687~nm.  The
    filled histogram refers to galaxies with confirmed spec-z's; the
    open histogram shows results based on photo-z's
    (P\'erez-Gonz\'alez et al.  2008). Shaded bands depict the
    expected redshifts for sources with emission in typical lines
    (e.g., H$\alpha$, [OII], Ly$\alpha$; some absorption features are
    also shown) lying in the F687W17 filter.}
  \label{fig:fig1}
\end{figure}

In Cava et al. (2012), we demonstrate that SHARDS data can be used to
segregate emission-line galaxies. That paper concentrates in the
selection performed with the F687W17 SHARDS filter (central
wavelength: 687~nm; width: 17~nm). Here, in Figure~\ref{fig:fig1}, we
present the redshift histogram for the emission-line galaxies selected
in Cava et al. (2012). More than 75\% of the spectroscopically
confirmed z$\sim$0.8 [OII] emitters in the GOODS-N field are selected
with our medium-band data.  The resemblance of the redshift
distribution built with photo-z's (open histogram) with that built
with spec-z's (filled) is remarkable, and the density peaks are placed
where one would expect a peak in the detection efficiency based on the
presence of an emission-line or absorption feature in the pass-band of
the chosen filter (F687W17, in this example).

\begin{figure}[!t]
\hspace{0.1cm}
  \includegraphics[width=7.8cm]{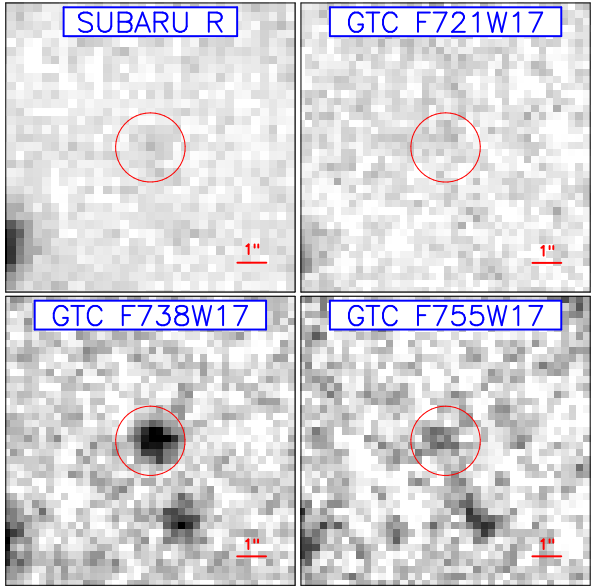}
  \caption{Postage stamps of a close-pair of Lyman-$\alpha$ emitters
    (LAEs) at z$\sim$5 in the ultra-deep data taken in the GOODS-N
    field by Subaru ($R$-band, resolution R$\sim$7) and GTC (SHARDS
    data, R$\sim$50).}
  \label{fig:fig2}
\end{figure}

Figure~\ref{fig:fig2} shows postage stamps for two Ly$\alpha$ emitter
candidates at z$\sim$5. The $R$-band data and the SHARDS images at
wavelengths shorter than 738~nm show blank fields, but the F738W17
filter reveals two bright objects whose Ly$\alpha$ emission is located
within the mentioned filter. According to our analysis of the whole
SED to obtain photo-z's for both sources, these two galaxies lie at
z$\sim$5.

\section{High-z Red \& Dead Sources in SHARDS}
\label{sec:rds}

\begin{figure}[!t]
  \includegraphics[width=7.9cm,angle=-90]{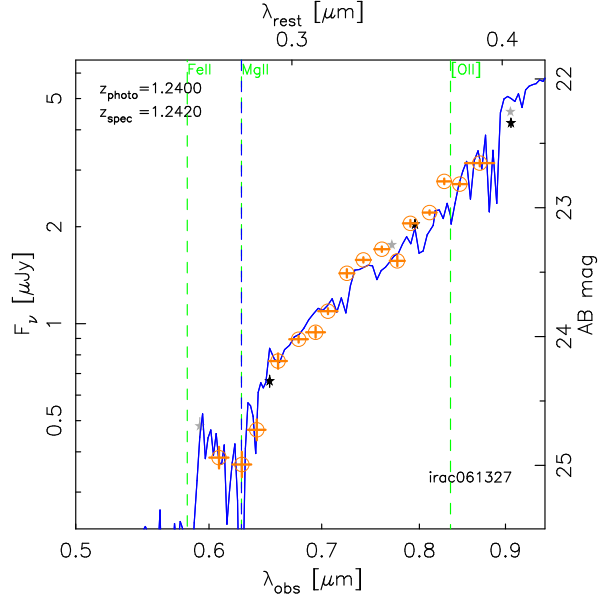}
  \caption{SED of J123704.36$+$621335.0, one of the spectroscopically
    confirmed compact massive spheroidal galaxies in GOODS-N
    (10$^{11.03\pm0.04}$~$\mathcal{M}_\sun$, P\'erez-Gonz\'alez et al.
    2008; see also Newman et al. 2010). The broad-band data (black and
    gray symbols) have been fitted to a stellar population model (in
    blue). SHARDS data (depicted in orange jointly with filter widths
    and errors), and the positions of the Mg absorption feature (also
    formed by FeII lines) and the [OII] emission-line are marked.}
  \label{fig:fig3}
\end{figure}

Figure~\ref{fig:fig3} shows the SED of one of the few
spectroscopically confirmed massive compact galaxies at z$>$1 in the
GOODS-N field (Newman et al. 2010). The SHARDS data allow to measure
absorption features such as the Mg$_\mathrm{UV}$ (around 280~nm
rest-frame), which can be used to constrain the stellar population
models and obtain a robust determination of the age. For this galaxy,
the SHARDS fluxes also reveal emission in the [OII] line, probably
coming from some residual star formation (the source also counts with
a faint detection in the MIPS 24~$\mu$m band). So the galaxy is not
completely dead, as the broad-band data implied, but maybe in a
post-starburst phase.




\begin{thebibliography}
\bibitem[foo(1000)]{bb}Arnouts, et al., 2007, A\&A, 476,137
\bibitem[foo(1000)]{bb}Baugh, et al., 1996, MNRAS, 283, 1361
\bibitem[foo(1000)]{bb}Buitrago, et al., 2008, ApJ, 687, 61
\bibitem[foo(1000)]{bb}Bundy, et al., 2007, MNRAS, 665, 5
\bibitem[foo(1000)]{bb}Cava, et al., 2012, this volume
\bibitem[foo(1000)]{bb}Cimatti, et al., 2006, A\&A, 453, 29
\bibitem[foo(1000)]{bb}Cole, et al., 2000, MNRAS, 319, 168
\bibitem[foo(1000)]{bb}Cowie, et al., 1996, AJ, 112, 839
\bibitem[foo(1000)]{bb}Croton, et al., 2006, MNRAS, 365, 11
\bibitem[foo(1000)]{bb}Daddi, et al., 2004, ApJ, 617, 746
\bibitem[foo(1000)]{bb}Daddi, et al., 2005, ApJ, 626, 680
\bibitem[foo(1000)]{bb}de Lucia, et al., 2006, MNRAS, 366, 499
\bibitem[foo(1000)]{bb}Fontana, et al., 2009, A\&A, 501, 15
\bibitem[foo(1000)]{bb}Franx, et al., 2003, ApJ, 587, 79
\bibitem[foo(1000)]{bb}Hopkins, et al., 2008, ApJS, 175, 390
\bibitem[foo(1000)]{bb}Hopkins, et al., 2009, MNRAS, 398, 898
\bibitem[foo(1000)]{bb}Ilbert, et al., 2009, ApJ, 709, 644
\bibitem[foo(1000)]{bb}Kriek, et al., 2008, ApJ, 682, 896
\bibitem[foo(1000)]{bb}Labb\'e, et al., 2007, ApJ, 665, 944
\bibitem[foo(1000)]{bb}Newman, et al., 2010, ApJ, 717, 103
\bibitem[foo(1000)]{bb}Ouchi, et al., 2008, ApJS, 176, 301
\bibitem[foo(1000)]{bb}P\'{e}rez Gonz\'{a}lez, et al., 2008, ApJ, 675, 234
\bibitem[foo(1000)]{bb}Rodr\'{\i}guez-Espinosa, et al., 2012, this volume
\bibitem[foo(1000)]{bb}Toft, et al., 2007, ApJ, 671, 285
\bibitem[foo(1000)]{bb}Trujillo, et al., 2007, MNRAS, 382, 109
\bibitem[foo(1000)]{bb}Villar, et al., 2008, ApJ, 677, 169
\end{thebibliography}
\end{document}